\documentclass[aps,pra,reprint,groupedaddress]{revtex4-1}
\usepackage{mathrsfs}
\usepackage{amsfonts}
\usepackage{graphicx}
\usepackage{}% Include figure files
\usepackage{amsmath}% Include figure files
\usepackage{dcolumn}% Align table columns on decimal point
\usepackage{color}
\usepackage[colorlinks, linkcolor=blue, citecolor=blue,hyperindex,bookmarks=false,pdfstartview=FitH]{hyperref}
\begin{document}

%\preprint{}

\title{Directional amplifier in an optomechanical system with optical gain}

\author{Cheng Jiang$^{1,2}$}
\author{L. N. Song$^{1}$}
\author{Yong Li$^{1,3}$}\email{liyong@csrc.ac.cn}

\affiliation{$^1$Beijing Computational Science Research Center, Beijing 100193, China}
\affiliation{$^2$School of Physics and Electronic Electrical Engineering, Huaiyin Normal University, 111 West Chang Jiang Road, Huai'an 223300, China}
\affiliation{$^3$Synergetic Innovation Center for Quantum Effects and Applications, Hunan Normal University, Changsha 410081, China}

\date{\today}

\begin{abstract}
Directional amplifiers are crucial nonreciprocal devices in both classical and quantum information processing. Here we propose a scheme for realizing a directional amplifier between optical and microwave fields based on an optomechanical system with optical gain, where an active optical cavity and two passive microwave cavities are, respectively, coupled to a common mechanical resonator via radiation pressure. The two passive cavities are coupled via hopping interaction to facilitate the directional amplification between the active and passive cavities. We obtain the condition of achieving optical directional amplification and find that the direction of amplification can be controlled by the phase differences between the effective optomechanical couplings. The effects of the gain rate of the active cavity and the effective coupling strengths on the maximum gain of the amplifier are discussed. We show that the noise added to this amplifier can be greatly suppressed in the large cooperativity limit.
\end{abstract}

%\keywords{}

\maketitle

\section{Introduction}
The field of cavity optomechanics studies the nonlinear interaction between the electromagnetic cavity and mechanical resonator via radiation pressure~\cite{Aspelmeyer1,Marquardt,XiongH}, where the effective optomechanical coupling strength can be greatly enhanced by driving the cavity with a strong pump field. By applying a pump field that is tuned to the lower motional sideband of the cavity field, optomechanics has witnessed great achievements such as ground-state cooling of the mechanical resonator~\cite{Chan,Teufel1}, optomechanically induced transparency~\cite{Agarwal1,Weis,Naeini1}, quantum state transfer~\cite{Tian1,WangYD1,Andrews}, and so on. Moreover, for the pump field that is tuned to the upper motional sideband of the cavity field, quantum entanglement~\cite{Tian3,WangYD3} and microwave amplification~\cite{Massel1,Korppi1,Korppi2,Toth} have been investigated.

On the other hand, nonreciprocal elements such as isolators, circulators, and directional amplifiers play a crucial role in communication and quantum information processing. To achieve nonreciprocity, one need to break the time-reversal symmetry inherent in the governing electromagnetic-wave equations in linear and nonmagnetic media~\cite{Potton}. The traditional method to break the time-reversal symmetry is based on the magneto-optical effects (e.g., Farady rotation)~\cite{Haldane,Khanikaev,Bi}, which have the disadvantage of being bulky, costly, and unsuitable for on-chip integration. Recently, several alternative effects have been used to implement nonreciprocal optical devices, including dynamic spatiotemporal modulation of the refractive-index~\cite{Lira,Fang1}, angular momentum biasing in photonic or acoustic systems~\cite{Fleury,Estep,WangDW}, and optical nonlinearity~\cite{FanL,Chang,GuoX}. Furthermore, reconfigurable Josephson circulator and directional amplifier have been demonstrated in superconducting microwave circuit by a set of parametric pumps~\cite{Sliwa,Lecocq1}.

More recently, radiation-pressure-induced parametric coupling between cavity and mechanical modes in optomechanics has been exploited to break the time-reversal symmetry, leading to an intensive research in nonreciprocity based on optomechanical coupling~\cite{Hafezi,XuXW1,Metelmann1,ShenZ,Tian2,Miri,Peterson}. Nonreciprocal devices including optomechanical isolators, circulators, and directional amplifiers have been proposed theoretically~\cite{XuXW2,Malz} and realized experimentally~\cite{Ruesink,Bernier,Barzanjeh,FangKJ,ShenZ2}. Many of these works rely on controlling the relative phases of the pumps applied to the cavity modes to achieve nonreciprocity. Metelmann and Clerk have proposed a reservoir engineering approach for nonreciprocal transmission and amplification by modulating the interaction between the system and the dissipative reservoir~\cite{Metelmann1}, and this approach has recently been applied to demonstrate the directional amplifier in an optomechanical crystal circuit~\cite{FangKJ}. Furthermore, it has been shown that directional amplification can be realized in a double-cavity optomechanical system with mechanical gain~\cite{ZhangXZ} or by introducing an additional mechanical drive~\cite{LiY}.

In this paper, we propose a scheme for realizing the directional amplifier between light fields with different frequencies (e.g. optical field and microwave field) in a triple-cavity optomechanical system with optical gain, where one cavity is doped with optical gain medium (i.e., active cavity) and the other two passive cavities are coupled to each other via hopping interaction, and the three single-mode cavities are coupled to a common mechanical resonator, respectively. This model is similar to that proposed in Ref.~\cite{Tian2}, where nonreciprocal quantum-state conversion between microwave and optical photons was studied without including any gain medium. By introducing the optical gain, nonreciprocity has recently been observed in parity-time-symmetric ($\mathcal {P}\mathcal {T}$-symmetric) microcavities with balanced gain and loss~\cite{Chang,PengB,LuoXB}. Subsequently, optomechanical systems with optical gain have witnessed rapid progress, including phonon laser~\cite{JingH1,HeB}, optomechanically induced transparency~\cite{JingH2} and all-optical photon transport switching~\cite{DuL}, $\mathcal {P}\mathcal {T}$-symmetry-breaking chaos~\cite{LuXY}, enhanced ground-state cooling of the mechanical resonator~\cite{LiuYL}, and enhanced sensitivity of detecting the mechanical motion~\cite{LiuZP}. Here we show that the optomechanical system with optical gain can operate as a directional amplifier between the active and passive cavities, where the direction of amplification can be controlled by adjusting the phase differences between the effective optomechanical couplings. Different from previous works~\cite{Ruesink,Malz,ShenZ2,LiY}, we find that the optical gain is the origin of amplification, instead of the blue-detuned optical pump fields~\cite{Ruesink,Malz,ShenZ2} or additional mechanical drive~\cite{LiY}. Furthermore, the mechanical noise can be greatly suppressed by increasing the cooperativity associated with the active cavity. It is worth pointing out that no direct coupling is needed between the active and passive cavities in this optomechanical system. Therefore, directional amplification can be realized between two cavities with vastly different frequencies. For example, the active cavity is an optical cavity and the two passive cavities can be microwave cavities. Finally, we briefly discuss the group delay of the amplified field and find that it can be prolonged evidently compared to the case without optical gain.

The remainder of the paper is organized as follows. In Sec.~\ref{MOD}, we introduce the theoretical mode and derive the transmission matrix between the input and output operators. In Sec.~\ref{DAM}, we obtain the optimal condition for directional amplifier and study in detail the effects of phase difference, optical gain rate, and the effective coupling strength. Gain, gain-bandwidth product, and added noise are also discussed. In Sec.~\ref{SLE}, we briefly investigate the group delay of the transmitted probe field. We finally summarize our work in Sec.~\ref{CONC}.

\section{Model}\label{MOD}
\begin{figure}
\centering
\includegraphics[width=8.5cm]{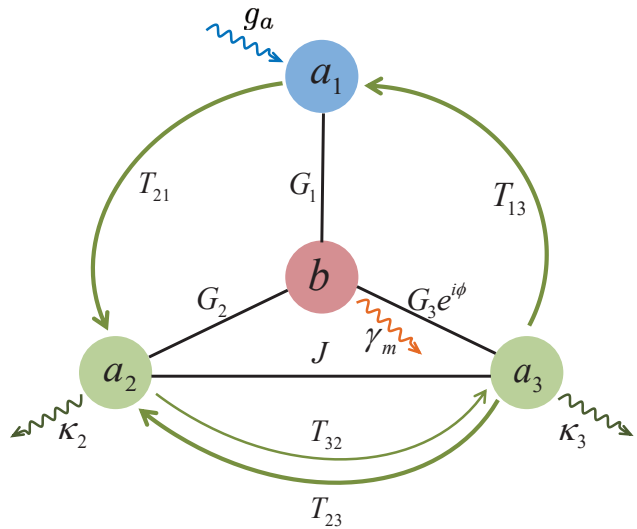}
\caption{Schematic diagram of the optomechanical system with optical gain. The active cavity $a_1$ with effective gain rate $g_a$ and the two passive cavities $a_2~(a_3)$ with decay rate $\kappa_2~(\kappa_3)$ are, respectively, coupled to a common mechanical resonator $b$ with damping rate $\gamma_m$. Cavities $a_2$ and $a_3$ are coupled via hopping interaction to facilitate the directional amplification. The solid arrows and $T_{ij}~(i,j=1,2,3)$ represent the transmission from the cavity $a_j$ to the cavity $a_i$.}
\label{Fig1}
\end{figure}
The optomechanical system under consideration is schematically shown in Fig.~\ref{Fig1}. Three cavity modes $a_1,a_2,a_3$ are coupled to a common mechanical mode $b$ via radiation pressure, respectively. Cavity $a_1$ can have vastly different frequency from that of cavities $a_2$ and $a_3$, e.g., $a_1$ can be an optical microcavity and $a_2$, $a_3$ are microwave cavities. Here we consider the cavity $a_1$ is an active cavity, which can be fabricated from $\mathrm{Er^{3+}-}$doped silica and can emit photons in the 1550-nm band by optically pumping $\mathrm{Er}^{3+}$ ions with a pump laser in the 1460-nm band~\cite{PengB} as an example. Furthermore, passive cavity $a_2$ is directly coupled to passive cavity $a_3$ via hopping interaction to facilitate the directional amplification between the active and passive cavities. We apply a strong driving field on each cavity mode to establish the parametric coupling. The Hamiltonian of this optomechanical system can be written as
\begin{eqnarray}
H=&&\sum_{k=1}^3\omega_k a_{k}^{\dagger}a_{k}+\sum_{k=1}^{3}g_{k}a_{k}^{\dagger}a_{k}(b^{\dagger}+b)+\omega_m b^{\dagger}b\nonumber\\
&&+J(a_2^{\dagger}a_3+a_3^{\dagger}a_2)\nonumber\\
&&+\sum_{k=1}^3\left(\varepsilon_{k}a_{k}e^{i\omega_{d,k}t}+\varepsilon_{k}^*
a_{k}^{\dagger}e^{-i\omega_{d,k}t}\right),
\end{eqnarray}
where $a_k$~$(a_k^{\dagger})$ is the annihilation (creation) operator of the cavity mode $a_k$~$(k=1,2,3)$ with resonance frequency $\omega_k$, $b$~$(b^{\dagger})$ is the annihilation (creation) operator of the mechanical mode $b$ with resonance frequency $\omega_m$, and $g_k$ is the single-photon optomechanical coupling strength between the cavity mode $k$ and the mechanical mode $b$. The fourth term represents the interaction between the cavities $a_2$ and $a_3$ with $J$ being the coupling strength. The last term describes the coupling between the cavity modes and the driving fields with amplitude $\varepsilon_k$ and frequency $\omega_{d,k}$. We can write each operator for the cavity modes as the sum of its classical mean value and quantum fluctuation operator, i.e., $a_k\rightarrow\alpha_k e^{-i\omega_{d,k}t}+a_k$, where the classical amplitude $\alpha_k$ is determined by solving the classical equation of motion~\cite{Barzanjeh}. Further moving to the rotating frame with respect to $H_0=\sum_{k=1}^{3}\omega_k a_k^{\dagger}a_k+\omega_m b^{\dagger}b$, and neglecting the counter-rotating and higher-order terms with $|\alpha_k|\gg1$, we can obtain the following linearized Hamiltonian
\begin{eqnarray}
H=&&G_1(a_1^{\dagger}b+a_1b^{\dagger})+G_2(a_2^{\dagger}b+a_2 b^{\dagger})\nonumber\\
&&+G_3(a_3^{\dagger}b e^{-i\phi}+a_3 b^{\dagger}e^{i\phi})+J(a_2^{\dagger}a_3+a_3^{\dagger}a_2),
\end{eqnarray}
where $G_k=g_k|\alpha_k|$ ($k=1,2,3$) is the effective coupling strength between the mechanical mode and the cavity $a_k$, and the phases of $\alpha_k$ have been absorbed by redefining the operators $a_k$ and $b$, and only the phase difference $\phi$ between them has physical effects. Here we have assumed that $\omega_2=\omega_3$, $\omega_{d,2}=\omega_{d,3}$, and $\Delta_k=\omega_k-\omega_{d,k}=\omega_m$.

According to the Heisenberg equations of motion and adding the corresponding damping and noise terms, we can get the following quantum Langevin equations (QLEs) \cite{Toth,Bernier,Barzanjeh}
\begin{eqnarray}
\dot{a_1}=&&-iG_1 b+\frac{g_a}{2}a_1+\sqrt{\kappa_{\mathrm{ex},1}}a_{1,\mathrm{in}}+\sqrt{\kappa_{0,1}}a_{1,\mathrm{in}}^{(0)}
\nonumber\\&&+\sqrt{g}a_{1,\mathrm{in}}^{(g)},\\\label{EQ3}
\dot{a_2}=&&-iG_2 b-iJ a_3-\frac{\kappa_2}{2} a_2+\sqrt{\kappa_{\mathrm{ex},2}}a_{2,\mathrm{in}}\nonumber\\&&+\sqrt{\kappa_{0,2}}
a_{2,\mathrm{in}}^{(0)},\\\label{EQ4}
\dot{a_3}=&&-iG_3 b e^{-i\phi}-iJ a_2-\frac{\kappa_3}{2}
a_3+\sqrt{\kappa_{\mathrm{ex},3}}a_{3,\mathrm{in}}\nonumber\\&&
+\sqrt{\kappa_{0,3}}a_{3,\mathrm{in}}^{(0)},\\\label{EQ5}
\dot{b}=&&-iG_1 a_1-iG_2 a_2-i G_3 a_3 e^{i\phi}-\frac{\gamma_m}{2} b+\sqrt{\gamma_m}b_{\mathrm{in}},\label{EQ6}
\end{eqnarray}
where $g_a=g-\kappa_1$ is the effective gain rate of cavity $a_1$ and $g$ is the gain which can be provided by pumping the $\mathrm{Er}^{+3}$ ions in cavity $a_1$ \cite{Chang,PengB}; the total decay rate of cavity $a_k~(k=1,2,3)$ is described by $\kappa_k=\kappa_{\mathrm{ex},k}+\kappa_{0,k}$, where $\kappa_{\mathrm{ex},k}$ and $\kappa_{0,k}$ are the external coupling rate and the intrinsic dissipation rates, respectively; $\gamma_m$ is the damping rate of the mechanical mode $b$. By introducing the Fourier transform of the operators
\begin{eqnarray}
o(\omega)=\int_{-\infty }^{+\infty}o(t)e^{i\omega t}dt,\\
o^{\dagger}(\omega)=\int_{-\infty }^{+\infty}o^{\dagger}(t)e^{i\omega t}dt,
\end{eqnarray} the zero-mean noise operators $a_{\mathrm{1,in}}^{(g)}$ and $a_{\mathrm{1,in}}^{(g){\dagger}}$ associated with the gain in cavity $a_1$ obey \cite{Agarwal,HeB,Kepesidis,LiuYL}
\begin{eqnarray}
&&\langle a_{\mathrm{1,in}}^{(g)}(\omega)a_{\mathrm{1,in}}^{(g){\dagger}}(\Omega)\rangle=0, \nonumber\\
&&\langle a_{\mathrm{1,in}}^{(g){\dagger}}(\Omega)a_{\mathrm{1,in}}^{(g)}(\omega)\rangle=2\pi\delta(\omega+\Omega).
\end{eqnarray}
Here we have assumed that the thermal photon occupation of cavity $a_1$ is zero because $\hbar\omega_1/k_{\mathrm{B}}T_e\gg1$ at optical frequencies, where $k_{\mathrm{B}}$ is the Boltzmann constant and $T_e$ is the temperature of the environment.

Furthermore, the input field $a_{k,\mathrm{in}}$ incident on the cavity $a_k~(k=1,2,3)$ via the external coupling satisfies the following correlation functions \cite{XuXW1,XuXW2,Agarwal4}
\begin{eqnarray}
&&\langle a_{k,\mathrm{in}}(\omega)a_{k,\mathrm{in}}^{\dagger}(\Omega)\rangle=2\pi[s_{k,\mathrm{in}}(\omega)+1]\delta(\omega+\Omega),
\nonumber\\
&&\langle a_{k,\mathrm{in}}^{\dagger}(\Omega) a_{k,\mathrm{in}}(\omega)\rangle=2\pi s_{k,\mathrm{in}}(\omega)\delta(\omega+\Omega),
\end{eqnarray}
where the term 1 results from the effect of vacuum noise and $s_{k,\mathrm{in}}(\omega)$ denotes the weak probe field incident on the cavity $a_k$ via the external coupling.
The noise operators $a_{k,\mathrm{in}}^{(0)}$ and $b_{\mathrm{in}}$ correspond to the intrinsic dissipation of cavity $a_k$ and the damping of the mechanical resonator, respectively. Under the white noise assumption, the above zero-mean noise operators satisfy the nonzero correlation functions given by \cite{Toth,Bernier,Barzanjeh}
\begin{eqnarray}
&&\langle a_{k,\mathrm{in}}^{(0)}(\omega)a_{k,\mathrm{in}}^{(0)\dagger}(\Omega)\rangle=2\pi\delta(\omega+\Omega), \nonumber\\
&&\langle b_{\mathrm{in}}(\omega)b_{\mathrm{in}}^{\dagger}(\Omega)\rangle=2\pi(n_m+1)\delta(\omega+\Omega),\nonumber\\
&&\langle b_{\mathrm{in}}^{\dagger}(\Omega)b_{\mathrm{in}}(\omega)\rangle=2\pi n_m\delta(\omega+\Omega).
\end{eqnarray}
Here the thermal photon occupations of the cavities are assumed to be zero under the condition that the reservoirs of the cavities are at sufficiently low temperature \cite{Aspelmeyer1,Malz,Bernier}, and the thermal phonon occupation of the mechanical resonator $n_m={1}/[\exp(\hbar\omega_m/k_{\mathrm{B}}T_e)-1]$.

For convenience, the QLEs (\ref{EQ3})-(\ref{EQ6}) can be written in the following matrix form
\begin{equation}
\dot{\mu}=M\mu+L\mu_{in},\label{MatrEQ}
\end{equation}
where the vector $\mu=(a_1,a_2,a_3,b)^{\mathrm{T}}$, $\mu_{\mathrm{in}}=(a_{1,\mathrm{in}},a_{2,\mathrm{in}},\\a_{3,\mathrm{in}},a_{1,\mathrm{in}}^{(0)},
a_{2,\mathrm{in}}^{(0)},a_{3,\mathrm{in}}^{(0)},a_{1,\mathrm{in}}^{(g)},b_{\mathrm{in}})^{\mathrm{T}}$ with $\mathrm{T}$ representing the transpose, the coefficient matrix
\begin{equation}
M=
\left(
\begin{array}{cccc}
g_a/2 & 0 & 0 & -iG_1\\
0 & -\kappa_2/2 & -iJ & -iG_2\\
0 & -iJ & -\kappa_3/2 & -iG_3 e^{-i\phi}\\
-iG_1 & -iG_2 & -iG_3e^{i\phi} & -\gamma_m/2\\
\end{array}
\right), \label{EQ10}
\end{equation}
\begin{equation}
L^{\mathrm{T}}=
\left(
\begin{array}{cccc}
\sqrt{\kappa_{\mathrm{ex},1}} & 0 & 0 & 0\\
0 & \sqrt{\kappa_{\mathrm{ex},2}} & 0 & 0\\
0 & 0 & \sqrt{\kappa_{\mathrm{ex},3}} & 0\\
\sqrt{\kappa_{0,1}} & 0 & 0 &0\\
0 & \sqrt{\kappa_{0,2}} & 0 &0\\
0 & 0 & \sqrt{\kappa_{0,3}} &0\\
\sqrt{g} & 0 & 0 & 0\\
0 & 0 & 0 & \sqrt{\gamma_m}\\
\end{array}
\right). \label{EQL}
\end{equation}
%{\color{red}
%\begin{equation}
%L=
%\left(
%\begin{array}{cccccccc}
%\sqrt{\kappa_{\mathrm{ex},1}} & 0 & 0 & \sqrt{\kappa_{0,1}} & 0 & 0 & \sqrt{g} & 0\\
%0 & \sqrt{\kappa_{\mathrm{ex},2}} & 0 & 0 & \sqrt{\kappa_{0,2}} & 0 & 0 & 0\\
%0 & 0 & \sqrt{\kappa_{\mathrm{ex},3}} & 0 & 0 & \sqrt{\kappa_{0,3}} & 0 & 0\\
%0 & 0 & 0 & 0 & 0 & 0 & 0 & \sqrt{\gamma_m}\\
%\end{array}
%\right), \label{EQL}
%\end{equation}}

The system is stable only if the real parts of all the eigenvalues of matrix $M$ are negative. The stability condition can be derived by applying the Routh-Hurwitz criterion~\cite{DeJesus,Gradshteyn}, whose general form is too cumbersome
to give here. However, we will check numerically the stability condition in the following and choose the parameters in the stable regime.
The solution to Eq. (\ref{MatrEQ}) in the frequency domain is
\begin{equation}
\mu(\omega)=-(M+i\omega I)^{-1}L\mu_{\mathrm{in}}(\omega),
\label{FreEQ}
\end{equation}
where $I$ represents the unitary matrix. Upon substituting Eq.~(\ref{FreEQ}) into the standard input-output relation $\mu_{\mathrm{out}}(\omega)=\mu_{\mathrm{in}}(\omega)-L^{\mathrm{T}}\mu(\omega)$, we can obtain
\begin{equation}
\mu_{\mathrm{out}}(\omega)=T(\omega)\mu_{\mathrm{in}}(\omega),
\end{equation}
where the output field vector $\mu_{\mathrm{out}}(\omega)$ is the Fourier transform of $\mu_{\mathrm{out}}=(a_{1,\mathrm{out}},a_{2,\mathrm{out}},a_{3,\mathrm{out}},a_{1,\mathrm{out}}^{(0)},
a_{2,\mathrm{out}}^{(0)},a_{3,\mathrm{out}}^{(0)},\\a_{1,\mathrm{out}}^{(g)},b_{\mathrm{out}})^{\mathrm{T}}$, and the transmission matrix is given by
\begin{equation}
T(\omega)=I+L^{\mathrm{T}}(M+i\omega I)^{-1}L.\label{TransMa}
\end{equation}
Here the matrix element $T_{ij}(\omega)$ ($i,j=1,2,3$) describes the transmission amplitude of the signal incident on the cavity $a_j$ and output from the cavity $a_i$ via the external coupling.

\section{Directional Amplifier}\label{DAM}
In this section, we consider how to realize the directional amplification between cavity modes $a_1$ and $a_2$ when an input probe field is resonant with the cavity frequency, i.e., $\omega=0$. According to Eqs.~(\ref{EQ10}) and (\ref{TransMa}), we can obtain the transmission matrix elements $T_{21}$ and $T_{12}$ as follows
\begin{eqnarray}
&&T_{12}(\omega)=-\frac{\sqrt{\eta_1\eta_2\kappa_1\kappa_2}}{A(\omega)}G_1(G_2\Gamma_3+iJG_3e^{i\phi}),\label{T12}\\
&&T_{21}(\omega)=-\frac{\sqrt{\eta_1\eta_2\kappa_1\kappa_2}}{A(\omega)}G_1(G_2\Gamma_3+iJG_3e^{-i\phi}),\label{T21}
\end{eqnarray}
where $A(\omega)=\Gamma_1(\Gamma_2\Gamma_3\Gamma_m+\Gamma_3 G_2^2+\Gamma_2 G_3^2+\Gamma_m J^2+2iG_2 G_3 J\mathrm{cos}\phi)+G_1^2(\Gamma_2\Gamma_3+J^2)$, $\Gamma_1=g_a/2+i\omega,\Gamma_2=-\kappa_2/2+i\omega,
\Gamma_3=-\kappa_3/2+i\omega$, $\Gamma_m=-\gamma_m/2+i\omega$, and $\eta_k=\kappa_{\mathrm{ex},k}/\kappa_k~(k=1,2,3)$ is the coupling efficiency for the cavity $a_k$ \cite{Barzanjeh,Peterson,Miri}.

In order to realize the directional amplifier, we require e.g. that the probe field input from cavity $a_1$ can be amplified when it is transmitted from cavity $a_2$, but the probe field input from cavity $a_2$ cannot be transmitted from cavity $a_1$, i.e., $|T_{21}|^2>1$ and $|T_{12}|^2=0$. We can get from Eqs.~(\ref{T12}-\ref{T21}) that $|T_{12}(0)|=0$ and $|T_{21}(0)|\neq0$ if $\phi=-\pi/2$ and $G_3=G_2\kappa_3/(2J)$, and we will show later that $|T_{21}|^2$ can be larger than 1 due to the optical gain of cavity $a_1$. In addition, to prevent loss of the input field to other modes such as $a_3$ and $b$, it is desirable that $|T_{i1}/T_{21}|\ll1$ $(i\neq2)$ when $|T_{12}|^2=0$. By choosing $J=\sqrt{\kappa_2\kappa_3}/2,$  we can obtain that $|T_{31}|=0$ and $|T_{41}/T_{21}|=\sqrt{\gamma_m\kappa_{\mathrm{ex,2}}}/{(2G_2)}\ll1.$ Therefore, the conditions of directional amplification from cavity $a_1$ to cavity $a_2$ for an incident probe field with $\omega=0$ include
\begin{eqnarray}
\phi=-\pi/2, \ \ G_3=G_2\kappa_3/(2J), \ \ J=\sqrt{\kappa_2\kappa_3}/2.
\end{eqnarray}
Under the above conditions, the transmission amplitude $T_{21}$ on resonance can be simplified as
\begin{eqnarray}
T_{21}(0)&=&\frac{8\sqrt{\eta_1\eta_2\kappa_1\kappa_2}G_1G_2}{4\kappa_2G_1^2-4g_aG_2^2-g_a\kappa_2\gamma_m}\nonumber\\
&=&\frac{2\sqrt{\eta_1\eta_2C_1C_2}\kappa_1/g_a}{C_1\kappa_1/g_a-C_2-1}\label{EQ20}
\end{eqnarray}
with the optomechanical cooperativity $C_k=4G_k^2/(\kappa_k\gamma_m)$ for $k=1,~2$. The effective gain rate $g_a$ can be controlled by tuning the gain rate $g$ and we can assume $g_a=\kappa_1$ for simplicity. The gain of the amplifier is then given by~\cite{Malz}
\begin{equation}
\mathcal {G}=|T_{21}(0)|^2=\frac{4\eta_{1}\eta_{2}C_1C_2}{(C_1-C_2-1)^2}.
\end{equation}

\begin{figure}
\centering
\includegraphics[width=8.5cm]{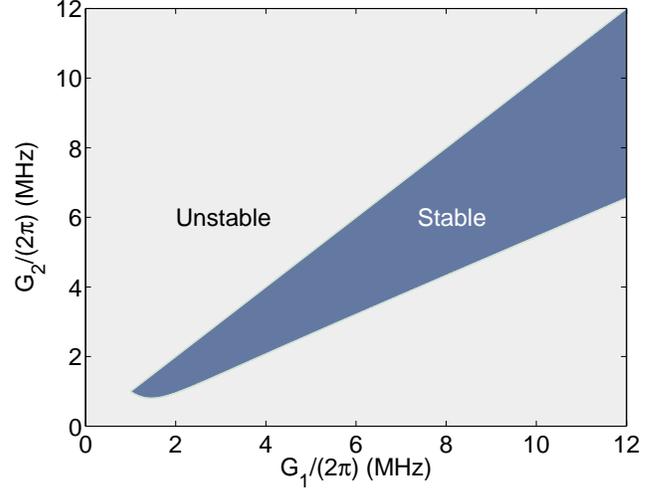}
\caption{Stability diagram with respect to $G_1$ and $G_2$. Other parameters are $g_a/2\pi=\kappa_{2}/2\pi=2$\,MHz, $\kappa_2/2\pi=2$\,MHz,
$\kappa_3/2\pi=3$\,MHz, $\gamma_m=\kappa_2/100$, $\phi=-\pi/2$, $G_3=G_2\kappa_3/(2J)$, and $J=\sqrt{\kappa_2\kappa_3}/2$.}
\label{Fig2}
\end{figure}
\begin{figure}
\centering
\includegraphics[width=8.5cm]{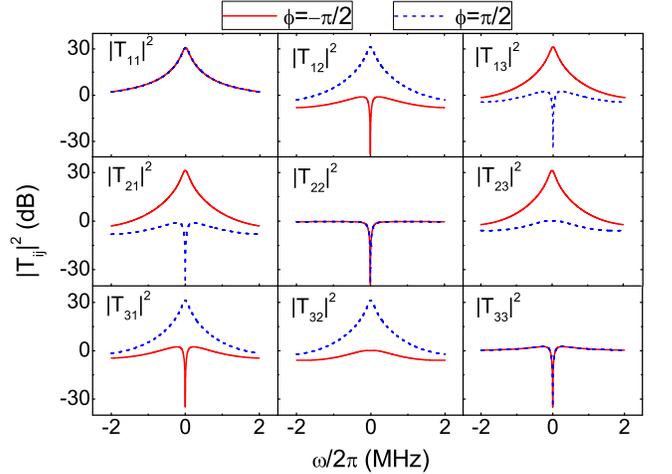}
\caption{Transmission probabilities $|T_{ij}|^2$ $(i,j=1,2,3)$ as functions of the probe detuning $\omega$ with $\phi=-\pi/2$ (red solid lines) or $\phi=\pi/2$ (blue dashed lines). Here $\eta_{1,2,3}=1$, $\kappa_{1}/2\pi=g_a/2\pi=2$ MHz, $G_1/2\pi=2$ MHz, and $\tilde{G}_2=\tilde{G}_1-0.1\sqrt{\tilde{G}_1}$, where $\tilde{G}_1$ $(\tilde{G}_2)$ is the dimensionless value of $G_1$ $(G_2)$ in units of MHz. For the other parameters, see Fig.~2.}
\label{Fig3}
\end{figure}
To realize the directional amplifier based on this optomechanical system with optical gain, the system should work in the stable regime. In Fig.~\ref{Fig2}, we plot the stability diagram with respect to the coupling strength $G_1$ and $G_2$, where the parameters are given as $g_a/2\pi=\kappa_1/2\pi=\kappa_2/2\pi=2$ MHz, $\kappa_3/2\pi=3$ MHz, $\gamma_m=\kappa_2/100$, $\phi=-\pi/2$, $G_3=G_2\kappa_3/(2J)$, and $J=\sqrt{\kappa_2\kappa_3}/2$. It can be seen that the system is stable only in a narrow regime due to the optical gain of cavity $a_1$. Furthermore, when the coupling strength $G_1$ is larger than a critical value ($G_1/2\pi \gtrsim 1.1$ MHz), the system can be stable as long as the coupling strength $G_2$ is a little smaller than $G_1$ because we have chosen $g_a=\kappa_2$. Therefore, we can choose $\tilde{G}_2=\tilde{G}_1-0.1\sqrt{\tilde{G}_1}$ for a given $G_1$ to ensure the system is stable, where $\tilde{G}_1$ $(\tilde{G}_2)$ is the dimensionless value of $G_1$ $(G_2)$ in units of MHz.

\begin{figure}
\centering
\includegraphics[width=8.5cm]{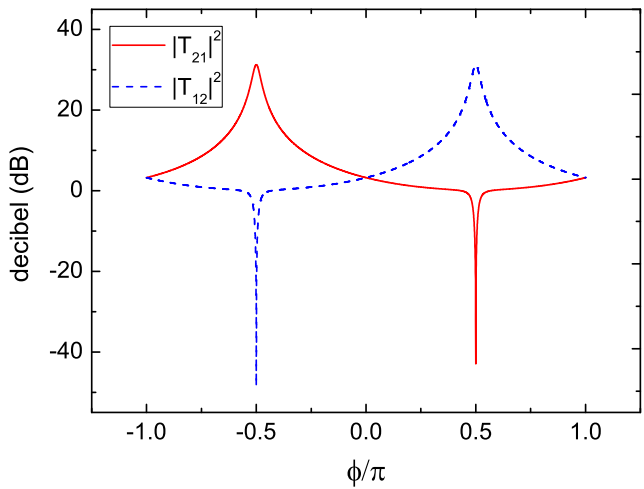}
\caption{(a) Transmission probabilities $|T_{12}|^2$ and $|T_{21}|^2$ versus the phase difference $\phi$ with $\omega=0$. The other parameters are the same as those in Fig.~\ref{Fig3}.}
\label{Fig4}
\end{figure}
According to the transmission matrix, we can study the frequency and phase dependence of the transmission probabilities.
Figure~\ref{Fig3} plots the transmission probabilities $|T_{ij}|^2$ $(i,j=1,2,3)$ between the cavities as a function of the probe detuning $\omega$ for phase difference $\phi=-\pi/2$ and $\phi=\pi/2$, respectively. When the phase difference is tuned to be $\phi=-\pi/2$, it can be seen that $|T_{21}|^2$ reaches the maximum value of about 30 dB (30 dB corresponds to $10^3$) but $|T_{12}|^2\approx0$ when $\omega=0$. Therefore, the signal incident on cavity $a_1$ can be greatly amplified when it is transmitted from cavity $a_2$, but the signal incident on cavity $a_2$ cannot be transmitted from cavity $a_1$. This directional amplification arises due to interference between two possible paths, where one path is along $a_1\rightarrow b\rightarrow a_2$, and the other path is along $a_1\rightarrow b\rightarrow a_3\rightarrow a_2$. When the phase difference $\phi=-\pi/2$, constructive interference between the two paths and the optical gain of cavity $a_1$ lead to the amplification for the optical transmission from cavity $a_1$ to cavity $a_2$, but the opposite direction is forbidden due to destructive interference ($|T_{12}|=0$). In addition, the signal input from cavity $a_3$ can be directionally amplified when it is transmitted from cavity $a_1$ with $|T_{13}|^2\approx30$ dB and $|T_{31}|^2\approx0$. Consequently, directional amplifier can be realized based on this optomechanical system, and the input signal can be amplified directionally along the route $a_3\rightarrow a_1\rightarrow a_2$. Furthermore, if we modulate the phase $\phi$ from $-\pi/2$ to $\pi/2$, the signal will be amplified along the opposite direction $a_2\rightarrow a_1\rightarrow a_3$, which can be illustrated by changing the labels from $T_{ij}$ to $T_{ji}$ and altering the directions of all the arrows in Fig.~\ref{Fig1}.

In what follows, we mainly consider the directional amplification between cavities $a_1$ and $a_2$. In Fig.~\ref{Fig4}, we plot the transmission probabilities $|T_{12}|^2$ and $|T_{21}|^2$ as a function of the phase difference $\phi$ for $\omega=0$. It can be seen that $|T_{12}|^2=|T_{21}|^2$ when $\phi=0$ and $\pm\pi$, therefore the Lorentz reciprocal theorem is satisfied and the response of this optomechanical system to the signal field is reciprocal. However, when $\phi\neq n\pi$ ($n$ is an integer), the time-reversal symmetry is broken, and the optomechanical system with optical gain exhibits a nonreciprocal response. In the regime $-\pi<\phi<0$, we have $|T_{12}|^2<|T_{21}|^2$, but $|T_{12}|^2>|T_{21}|^2$ when $0<\phi<\pi$. The optimal nonreciprocal response is obtained as $\phi=-\pi/2$ [$|T_{21}|^2\approx30$ dB and $|T_{12}|^2\approx0$] and $\phi=\pi/2$ [$|T_{12}|^2\approx30$ dB and $|T_{21}|^2\approx0$] for the given parameters. Therefore, directional amplifier can be realized by modulating the phase difference $\phi$.

\begin{figure}
\centering
\includegraphics[width=8.5cm]{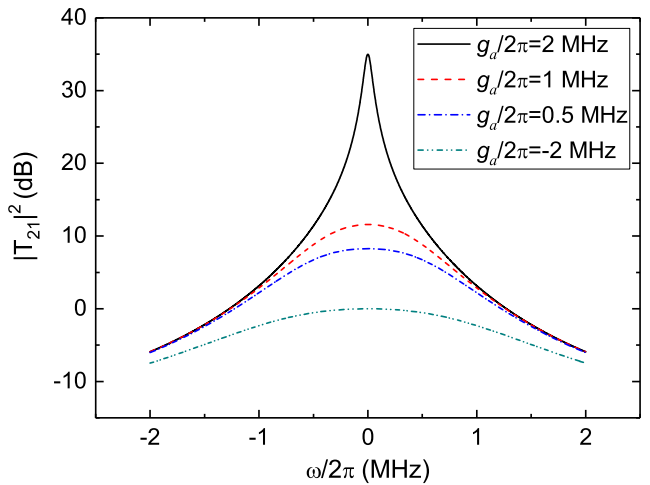}
\caption{Transmission probability $|T_{21}|^2$ as a function of the probe detuning $\omega$ for different values of effective gain rate $g_a$, where $g_a/2\pi=-2$ MHz means that cavity $a_1$ becomes the passive cavity. The other parameters are the same as those in Fig.~\ref{Fig3} except $\phi=-\pi/2$, $G_1/2\pi=5$ MHz and $\tilde{G}_2=\tilde{G}_1-0.1\sqrt{\tilde{G}_1}$, where $\tilde{G}_1$ $(\tilde{G}_2)$ is the dimensionless value of $G_1$ $(G_2)$ in units of MHz.}
\label{Fig5}
\end{figure}
Next, we study the effect of the effective gain rate $g_a$ of cavity $a_1$ on the transmission probability $|T_{21}|^2$. Figure \ref{Fig5} plots the transmission probability $|T_{21}|^2$ as a function of the probe detuning $\omega$ for different value of $\kappa_1$. When $g_a/2\pi$ decreases from 2\,MHz to 0.5\,MHz, the maximum transmission probability $|T_{21}|^2$ on resonance decreases from about 35\,dB to 8\,dB with $|T_{12}|=0$. Furthermore, if the active cavity $a_1$ becomes a passive cavity, i.e., $g_a/2\pi=-2$\,MHz, the transmission probability on resonance $|T_{21}|^2\approx1$, which indicates the appearance of optomechanically induced transparency~\cite{Agarwal1,Weis}. In this case, nonreciprocal transmission can still exist in this optomechanical system~\cite{Tian2}, but the input probe field cannot be amplified. Therefore, the optical gain is the origin of amplification, and the phase difference is responsible for the nonreciprocal transmission in this optomechanical system.

\subsection{Bandwidth and Gain-Bandwidth Product}

For the phase-preserving linear amplifier, the bandwidth generally decreases with the increase of the gain, which can be seen from Fig.~\ref{Fig5}. The bandwidth of the amplifier can be approximately obtained according to the denominator of $T_{21}(\omega)$ in Eq.~(\ref{T21}). If we assume $g_a=\kappa_2=\kappa_3=\kappa$, then $A(\omega)\approx\frac{1}{8}\kappa^3\gamma_m(C_1-C_2-1)+\frac{1}{4}\gamma_m\kappa^2(\kappa/\gamma_m-C_1)i\omega,$
where we only keep the terms to the first order of $\omega$ \cite{Malz}. The bandwidth $\Gamma$ is approximated by the smallest $|\omega|$ at which $2|A(0)|^2=|A(\omega)|^2$, and it can be given by
\begin{eqnarray}
\Gamma=\left|\frac{\kappa(C_1-C_2-1)}{\kappa/\gamma_m-C_1}\right|.
\end{eqnarray}
In this optomechanical system with optical gain, we can see from Fig.~\ref{Fig2} that the system is unstable when the cooperativity is small. Therefore, we consider the large cooperativity limit, i.e. $C_1>C_2\gg\kappa/\gamma_m$, then the bandwidth approaches $\Gamma=\kappa(C_1-C_2-1)/C_1,$ resulting in the gain-bandwidth product
$P\equiv\Gamma\sqrt{\mathcal {G}}\rightarrow2\kappa.$

\subsection{Added Noise}
The added number of noise quanta of the amplifier can be obtained by calculating the output spectra of cavity $a_2$, which is given by \cite{Malz,Metelmann}
\begin{eqnarray}
S_{2,\mathrm{out}}(\omega)&=&\frac{1}{2}\int\frac{d\Omega}{2\pi}
\langle a_{2,\mathrm{out}}(\omega)a_{2,\mathrm{out}}^{\dagger}(\Omega)\nonumber\\&&+
a_{2,\mathrm{out}}^{\dagger}(\Omega)a_{2,\mathrm{out}}(\omega)\rangle\nonumber\\
&=&\sum_{i=1}^3[s_{i,\mathrm{in}}(\omega)+\frac{1}{2}]|T_{2i}(\omega)|^2+\frac{1}{2}\sum_{i=4}^7|T_{2i}(\omega)|^2
\nonumber\\&&+(n_m+\frac{1}{2})|T_{28}(\omega)|^2,\label{EQ24}
\end{eqnarray}
where we have used the noise correlation function in the frequency domain and the relation $o^{\dagger}(\omega)=[o(-\omega)]^{\dagger}$. It can be seen from Eq. (\ref{EQ24}) that output spectra of cavity $a_2$ contains eight components. If we consider the directional amplification of the signal incident on the cavity $a_1$ and output from the cavity $a_2$ via the external coupling, i.e., $|T_{21}|^2$, then other transmission probabilities $|T_{2i}|^2$ $(i=2,3,4,...,8)$ associated with thermal occupation $n_i$ can be treated as noise and we can assume $s_{2,\mathrm{in}}=s_{3,\mathrm{in}}=0$. Therefore, the noise added to the amplifier is defined as
$\mathcal {N}_{2}(\omega)=\mathcal {G}^{-1}\sum_{i=2}^8(n_i+1/2)|T_{2i}(\omega)|^2$ \cite{CavesPRD,ClerkRMP,Nunnenkamp,Malz}.
The general form of $\mathcal {N}_{2}(\omega)$ is too cumbersome to give here. However, if $\eta_{1,2,3}=1$, $g_a=\kappa_1$, and $\omega=0$, the noise added to the output of cavity $a_2$ can be given by
\begin{equation}
\mathcal {N}_{2}(0)=\frac{1}{2}\frac{(C_1+C_2-1)^2}{4C_1C_2}+\left(n_m+\frac{1}{2}\right)\frac{1}{C_1}
+1,\label{EQ25}
\end{equation}
where we have assumed that the thermal photon occupations $n_i=0~(i=1,2,3,...,7)$ for the cavity modes and the thermal phonon occupation $n_8=n_m$ for the mechanical mode. The last term 1 on the right side of Eq. (\ref{EQ25}) results from the gain of cavity $a_1$. We can see from Eq. (\ref{EQ25}) that
the thermal noise from the mechanical resonator can be suppressed by increasing the cooperativity $C_1$. For large $C_1$ and $C_2$ with $C_1$ a little larger than $C_2$, we find at zero frequency $\mathcal {N}_{2}(0)\rightarrow1.5$.

\begin{figure}
\centering
\includegraphics[width=8.5cm]{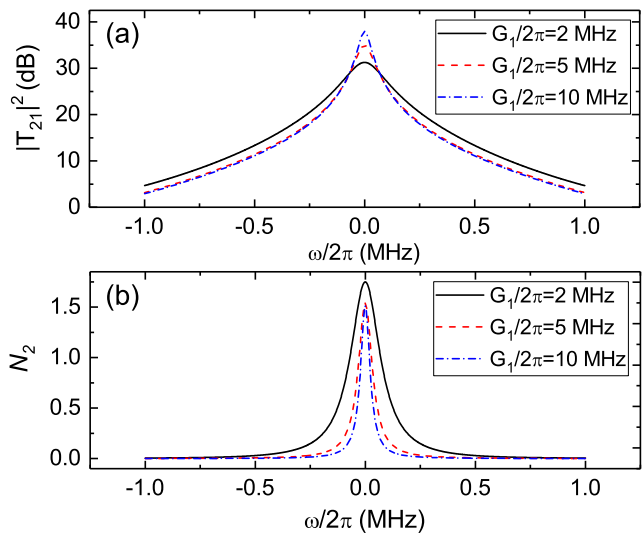}
\caption{(a) Transmission probability $|T_{21}|^2$ and (b) added noise $\mathcal {N}_{2}$ as a function of probe detuning $\omega$ with $G_1/2\pi=2,5,10$\,MHz and $\tilde{G}_2=\tilde{G}_1-0.1\sqrt{\tilde{G}_1}$, where $\tilde{G}_1$ $(\tilde{G}_2)$ is the dimensionless value of $G_1$ $(G_2)$ in units of MHz. The other parameters are the same as those in Fig.~\ref{Fig3} except $\phi=-\pi/2$ and} $n_m=100$.
\label{Fig6}
\end{figure}
Figures \ref{Fig6}(a) and \ref{Fig6}(b) plot respectively the transmission probability $|T_{21}|^2$ and noise added to the cavity $a_2$ with respect to the probe detuning $\omega$ for different coupling strengths $G_1$. It can be seen from Fig.~\ref{Fig6}(a) that the peak value of $|T_{21}|^2$ becomes larger when $G_1/2\pi$ increases from $2$ to $10$ MHz. Furthermore, Fig. \ref{Fig6}(b) shows that the added number of noise quanta $\mathcal {N}_{2}(\omega)$ decreases with increasing the optomechanical coupling constants $G_1$ and $G_2$. In particular, when $G_{1,2}$ are large enough, thus $C_{1,2}\gg 1$ and $\mathcal {N}_{2}(\omega)$ on resonance can approach to $1.5$.

\begin{figure}
\centering
\includegraphics[width=8.5cm]{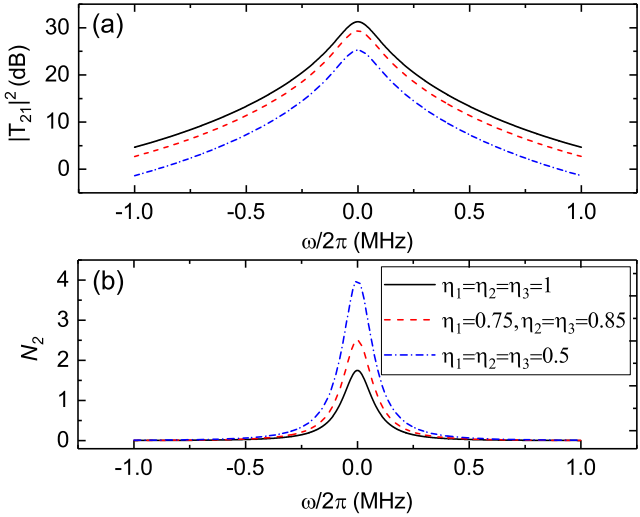}
\caption{(a) Transmission probability $|T_{21}|^2$ and (b) added noise $\mathcal {N}_{2}$ as a function of probe detuning $\omega$ for different values of coupling efficiency $\eta_k$ with $G_1/2\pi=2$ MHz and $\tilde{G}_2=\tilde{G}_1-0.1\sqrt{\tilde{G}_1}$, where $\tilde{G}_1$ $(\tilde{G}_2)$ is the dimensionless value of $G_1$ $(G_2)$ in units of MHz. The other parameters are the same as those in Fig.~\ref{Fig6}.}
\label{Fig7}
\end{figure}
Furthermore, the influence of internal dissipation rate of the cavity on the gain and added noise of the amplifier is discussed in Fig. \ref{Fig7}. We find that when the coupling efficiency $\eta_k$ reduced from 1 to 0.5 (critical coupling), the gain of the amplifier decreases, which can also be seen from Eq. (\ref{EQ20}). Meanwhile, Fig.~\ref{Fig7}(b) shows that the noise added to this amplifier becomes larger when the coupling efficiency reduces. Therefore, high coupling efficiency $\eta_k$ of each cavity is beneficial for enhancing the gain of the amplifier and suppressing the added noise.

\begin{figure}
\centering
\includegraphics[width=8.5cm]{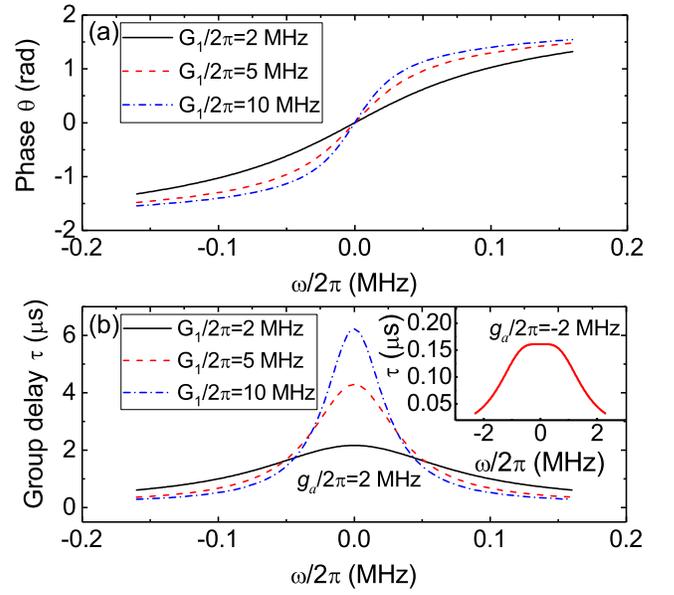}
\caption{(a) Phase and (b) group delay versus the probe detuning $\omega$ for different values of coupling strength $G_1$ with $g_a/2\pi=2$ MHz. The inset of Fig. 7(b) is the group delay $\tau$ versus the probe detuning $\omega$ when $g_a/2\pi=-2$ MHz and $G_1/2\pi=5$ MHz. The other parameters are the same as those in Fig.~\ref{Fig5}.}
\label{Fig8}
\end{figure}
\section{Slow light effect}\label{SLE}
Finally, we study the slow light effect in the transmitted probe field. It is well known that the probe field within the electromagnetically induced transparency (EIT) window usually suffers a rapid phase dispersion, which can lead to the dramatic reduction in its group velocity. In optomechanical systems, this kind of slow light effects has been extensively investigated in the past decade \cite{Weis,Naeini1}. Here we focus on the slow light effect associated with directional amplification. The optical group delay of the transmitted light is defined as \cite{Weis}
\begin{equation}
\tau=\frac{d\theta}{d\omega},
\end{equation}
where $\theta=\mathrm{arg}[T_{21}(\omega)]$ is the phase of the output field from cavity $a_2$ at the frequency of the probe field incident on cavity $a_1$.

Figure~\ref{Fig8} plots (a) phase and (b) group delay of the transmitted probe field from cavity $a_2$ for $G_1/2\pi=2,5,10$ MHz, respectively. We can see from Fig. \ref{Fig8}(a) that the amplified transmitted probe field is accompanied by rapid phase dispersion, and the slope of the phase dispersion around $\omega=0$ becomes larger when the coupling strength $G_1$ increases. Therefore, the maximum group delay is getting larger with increasing $G_1$, as can be seen from Fig. \ref{Fig8}(b). Furthermore, the inset of Fig. \ref{Fig8}(b) shows the group delay $\tau$ with respect to the probe detuning $\omega$ when cavity $a_1$ is also the passive cavity. In this case, optomechanically induced transparency can occur in the probe field transmitted from cavity $a_2$, as shown in Fig. \ref{Fig5}. When $G_1/2\pi=5$ MHz, we can see that the maximum group delay on resonance for the active cavity ($g_a/2\pi=2$ MHz) is about 4.3\,$\mu$s, while the maximum group delay for the passive cavity ($g_a/2\pi=-2$ MHz) is about 0.16\,$\mu$s. Therefore, the optomechanical system with optical gain allows for directional amplification with prolonged group delay.

\section{Conclusion}\label{CONC}
In summary, we have investigated the directional amplifier in a hybrid optomechanical system with optical gain. The transmission between the active cavity and the passive cavity can be directionally amplified, and the direction of the amplification depends on the phase difference between the effective optomechanical couplings. The maximum amplifier gain can be enhanced by increasing the gain rate of the active cavity and the effective coupling strengths. In the large cooperativity limit, the effect of the mechanical noise can be significantly suppressed in the cavity outputs. Furthermore, the group delay of the transmitted probe field in this optomechanical system with optical gain can be improved up to one or two orders of magnitude compared to the optomechanical system without optical gain.

\section*{ACKNOWLEDGMENTS}
This work is supported by the Science Challenge Project (under Grant No.~TZ2018003), the National Key R\&D Program of China grant 2016YFA0301200, the National Basic Research Program of China (under Grant No.~2014CB921403), and the National Natural Science Foundation of China (under Grants No.~11774024, No.~11534002, No.~U1530401). C.J. is supported by the Natural Science Foundation of China (NSFC) under Grant No.~11304110, the Postdoctoral Science Foundation of China under Grant No.~2017M620593, and Qing Lan Project of Universities in Jiangsu Province.

\end{document}